\newcommand{\beq}{\begin{equation}}
\newcommand{\eeq}{\end{equation}}
\newcommand{\bea}{\begin{eqnarray}}
\newcommand{\eea}{\end{eqnarray}}

\newcommand{\gsim}{\lower.7ex\hbox{$\;\stackrel{\textstyle>}{\sim}\;$}}
\newcommand{\lsim}{\lower.7ex\hbox{$\;\stackrel{\textstyle<}{\sim}\;$}}

\documentclass[aps,prev,preprintnumbers,floatfix,nofootinbib,11pt]{revtex4-1}
\usepackage{graphicx}
\usepackage{epstopdf}
\usepackage{mathrsfs}
\usepackage{amssymb}
\usepackage{verbatim}
\usepackage{color}
\usepackage{multirow}
\usepackage{amsmath}
\usepackage{float}
\usepackage[normalem]{ulem}
\usepackage[lofdepth,lotdepth,caption=false]{subfig}

\def\stacksymbols #1#2#3#4{\def\theguybelow{#2}
    \def\vp{\lower#3pt}
    \def\sp{\baselineskip0pt\lineskip#4pt}
    \mathrel{\mathpalette\intermediary#1}}

\def\intermediary#1#2{\vp\vbox{\sp
     \everycr={}\tabskip0pt
     \halign{$\mathsurround0pt#1\hfil##\hfil$\crcr#2\crcr
              \theguybelow\crcr}}}

\def\be{\begin{equation}}
\def\ee{\end{equation}}
\def\bea{\begin{eqnarray}}
\def\eea{\end{eqnarray}}
\begin{document}

\vspace*{1mm}

\title{On the Tau flavor of the cosmic neutrino flux }

\author{Yasaman Farzan$^{a}$}
\email{yasaman@theory.ipm.ac.ir}

\vspace{0.2cm}

\affiliation{
%${}^a$Departamento de F\'isica, Pontif\'icia Universidade Cat\'olica do Rio de Janeiro, Rio de Janeiro 22452-970, Brazil\\
${}^a$School of physics, Institute for Research in Fundamental Sciences (IPM),\\
P.O. Box 19395-5531, Tehran, Iran
}

\begin{abstract}  
	
	 Observation of high energy cosmic neutrinos by ICECUBE has ushered in a new era in exploring both cosmos and new physics beyond the Standard Model (SM). In the standard picture, although mostly $\nu_\mu$ and $\nu_e$ are produced in the source, oscillation will  produce $\nu_\tau$ {\it en route}. Certain beyond SM scenarios, like interaction with ultralight DM can alter this picture. Thus, the flavor composition of the cosmic neutrino flux can open up the possibility of exploring  certain beyond the SM scenarios that are inaccessible otherwise. We show that the $\tau$ flavor holds a special place among the neutrino flavors in elucidating new physics. Interpreting the two anomalous events observed by ANITA as $\nu_\tau$ events makes the tau flavor even more intriguing. We study how the detection of the two tau events by ICECUBE constrains the interaction of the neutrinos with ultralight dark matter and discuss the  implications of this interaction for even higher energy cosmic neutrinos detectable by future radio telescopes such as ARA, ARIANNA and GRAND. We also revisit the $3+1$ neutrino scheme as a solution to the two anomalous ANITA events and clarify a misconception that exists in the literature about the evolution of high energy neutrinos  in matter within the  $3+1$ scheme with a possibility of scattering off  nuclei.
	 We show that the existing bounds on the flux of $\nu_\tau$ with energy of EeV rules out this solution for the ANITA events. We show that the $3+1$ solution can be saved from both this bound and  from the  bound on the extra relativistic degrees of freedom in the early universe  by turning on the interaction of neutrinos with ultralight dark matter.
\end{abstract}

\maketitle

%%%%%%%%%%%%%%%%%%%%%%%%
%%%%%%%%%%%%%%%%%%%%%%%%
\section{Introduction}
%%%%%%%%%%%%%%%%%%%%%%%%
%%%%%%%%%%%%%%%%%%%%%%%%

For millennia, humans have gazed at the night sky with fascination and have tried to learn about  the stars.  In the 20th century, for the first time, humans were able to 
study the cosmos by electromagnetic waves outside the narrow optical window as
 well as by detecting the cosmic ray. In the second decade of the 21st century, we have truly entered the multi-messenger era as detecting  cosmic neutrinos  and gravitational waves provide alternative means to extract information about the faraway sources. ICECUBE with registering   over 60 events with energy higher than 60 TeV is a key player (see table II.3 of \cite{Abbasi:2020jmh}).
 The energy spectrum, directionality and the flavor of the incoming neutrinos can provide information on the potential sources and/or on beyond SM effects showing up at extremely large baselines and/or energies  associated with the cosmic rays. 
 
 The canonical scenario for the production of ultrahigh energy neutrinos is the scattering of high energy protons off other protons or photons leading to the charged pion production and their subsequent decay into neutrinos. The flavor composition of neutrinos at the source in this scenario is $\nu_e:\nu_\mu:\nu_\tau=1:2:0$ which due to the neutrino oscillation en route will be converted to $\nu_e:\nu_\mu:\nu_\tau=1:1:1$ when they reach the Earth. It has been shown that within the SM, deviation from the democratic prediction  ({\it i.e.,} $\nu_e :\nu_\mu:\nu_\tau=1:1:1$  in the Earth) cannot be significant. More precisely, only a small fraction of the area of the flavor ternary triangle is accessible by variation within the standard model ({\it i.e.,} by assuming total or partial muon energy loss at the source or by invoking a significant contribution from the neutron decay at the source) \cite{Bustamante:2015waa}. Even by turning on new physics such as the neutrino decay \cite{Denton:2018aml}, quantum decoherence, pseudo-Dirac neutrino, the accessible region in the ternary triangle remains small \cite{Bustamante:2015waa}.  In particular, the flux at Earth will still include a significant  $\nu_\tau$ component. \footnote{If the flux originates from the neutron decay, the flux at the source will be composed of $\nu_e:\nu_\mu:\nu_\tau=1:0:0$ which after propagating cosmic distances will have flavor composition of 
 	$\nu_e:\nu_\mu:\nu_\tau=0.58:0.21:0.21$.}
 
 In  Ref. \cite{Farzan:2018pnk}, it is shown that a coupling of current-current form  between neutrinos and 
the background ultralight dark matter can induce an effective mass  for neutrinos of Lorentz and flavor structure $V_\alpha \nu_\alpha^\dagger \nu_\alpha$. For conventional neutrino fluxes and beams such as those produced in the stars or core collapse supernovae, in the atmosphere or in the anthropogenic sources, this induced mass is negligible compared to $\Delta m^2_{12} /(2E_\nu)$ and its effects are therefore negligible. However, for higher energy cosmic neutrinos detectable by neutrino telescopes, the effective mass induced by the DM can be dominant.  If this effective mass is flavor diagonal (similarly to the prediction of the $L_\mu-L_\tau$ model proposed in \cite{Farzan:2018pnk}), the effective mixing  is suppressed in the environments such as the galactic Dark Matter (DM) halo where the  DM density is high.
As a result, a neutrino state produced in a halo maintains its flavor traveling from the DM halo of a far galaxy up to the halo of our galaxy. In other words, the flavor ratio would remain $\nu_e:\nu_\mu:\nu_\tau=1:2:0$. Recently, ICECUBE has observed two $\tau$ neutrino events with total deposited energies of 100 TeV    and  2 PeV  \cite{Abbasi:2020zmr}, shifting the observed flavor pattern towards the SM prediction and ruling out the zero $\nu_\tau$ flux hypothesis at 2.8 $\sigma$ C.L. In this paper, we derive bounds on the parameters of the model considering this new observation and discuss that the  dark matter effects proposed in \cite{Farzan:2018pnk} can still be significant for high energy cosmic neutrinos in the energy range of EeV.
We discuss the implications of different possible observations via detectors such as ARA \cite{ARA}, ARIANNA \cite{ARIANNA} and GRAND \cite{Fang:2017mhl}, considering that GRAND will only detect extensive showers from $\nu_\tau$ flux but the other two can distinguish $\nu_\tau$ events from Askaryan emission by  other flavors.

On the other hand, the ANITA radio telescope array flying over Antarctica  has reported two mysterious events that may originate from $\nu_\tau$ emerging from deep down the Earth with energy of $\sim$0.6 EeV. What makes these events anomalous is that the Earth is opaque for neutrinos at these energies. Although some explanations within the SM have been proposed \cite{deVries:2019gzs,Shoemaker:2019xlt}, the possibility of a glimpse on new physics by ANITA is too exciting to dismiss. One possibility which was first proposed in  \cite{Cherry:2018rxj} (see also \cite{Esmaili:2019pcy}) is based on the introduction of a new fermion originating from cosmologically distant sources with mean free path in the Earth mantle comparable to the chords corresponding to the ANITA events ({\it i.e.,} 5000-7500 km), converting to $\nu_\tau$ via scattering off matter in the vicinity of ANITA. An example for such messenger is the sterile neutrino mixed with $\nu_\tau$. That is the $3+1$ scheme with $U_{\tau 4}\ne 0$. After clarifying a misconception that exists in the literature about propagation of mass eigenstates in matter in the presence of absorption of the active component, we discuss the possible sources of production of such high energy sterile neutrinos. We show that, along with a $\nu_s$ flux, the source will also emit $\nu_\tau$ on whose flux at these energies there is a strong bound \cite{Safa,Pizzuto:2019vyj}.

Moreover, the $3+1$ scheme is ruled out by the cosmological bounds on the extra relativistic degrees of freedom in the early universe. As shown in \cite{Farzan:2019yvo}, the coupling of neutrinos to ultralight dark matter can save the $3+1$ scheme from the cosmological bounds as the effective induced flavor diagonal mass in the early universe will be large enough to suppress the effective mixing and therefore the $\nu_s$ production in the early universe.
We revisit the $3+1$ solution to the ANITA events in the presence of DM effects and show that the solution can be revived because the accompanying $\nu_\tau$ flux that reach the Earth can be small.

This paper is organized as follows. In sect. \ref{TAUproduction}, the possibility of $\nu_\tau$ production as a sub-dominant component in cosmic neutrino sources is briefly discussed.  In sect. \ref{anita+3+1}, the $3+1$ scheme as a solution for the ANITA anomaly is revisited and the predictions for the successor of ANITA are discussed.
 In sect. \ref{UDM}, a bound from recent $\nu_\tau$ observation by ICECUBE on the coupling of neutrinos to ultralight DM is derived. The implications for cosmogenic neutrinos and for other possible EeV neutrinos are also discussed. In sect. \ref{Shole}, we show that the coupling of neutrino to ultralight DM can relax the bounds on the 3+1 scheme. Conclusions are reviewed in sect. \ref{Fin}.
%%%%%%%%%%%%%%%%%%%%%%%%%%%%% 
%%%%%%%%%%%%%%%%%%%%%%%%%%%%%%
\section{$\nu_\tau$ production at the source \label{TAUproduction}}
 The high energy neutrinos can be produced by the scattering of accelerated high energy protons off either the protons in the source or off a photon gas.
 In the case of cosmogenic neutrinos, the scattering is off the ambient CMB photons. In the majority of the models for the neutrino production in the Tidal Disruption Events (TDE)  \cite{Wang:2011ip} (see also, \cite{Murase:2008zzc,Farrar:2008ex}), in GRBs \cite{Piran:2004ba} and in  the AGNs \cite{Nellen:1992dw,Stecker:1991vm}, the scattering of the energetic protons takes place off a distribution of photons rather than off protons. However, models for neutrinos from proton proton scattering in the blazars exist \cite{Sahakyan:2018voh,Liu:2018utd}. 
 Indeed, it seems the proton proton scattering explains the absence of a significant electromagnetic activity during the so-called ``historic neutrino flare" observed in the direction of blazar TXS 0506+056 during October 2014 to March 2015 \cite{Rodrigues:2018tku}.
 It is quite conceivable that for various source and even during different periods of activity, either $pp$ or $p\gamma$ scattering dominates. A systematic evaluation of the efficiency of the different possibilities at the source can be found in \cite{Murase:2018iyl}.  
 In the following, we discuss the production of $\nu_\tau$ at the source for both of these mechanisms.
 
  The production of neutrinos from the scattering of protons off photons becomes efficient when the center of mass energy of the proton photon system is equal to the $\Delta$ mass (1.2 GeV) leading to $p^+ +\gamma \to \Delta^+ \to n +\pi^+$. This means $E_\gamma E_p (1+\cos \theta)=(m_\Delta^2-m_p^2)/2=0.28 ~{\rm GeV}^2$ where $\theta$ is the angle between the momenta of the proton and photon. The neutrinos from the resonance $\Delta$ production are only $\nu_\mu$, $\nu_e$ and $\bar{\nu}_e$. To produce $\nu_\tau$, the $c$ quarks have to be produced. 
  The process must conserve the baryon number. Moreover, the strong and electromagnetic interactions respect $c$-number conservation. Thus, either $c\bar{c}$ should be produced in pairs or the photon should interact with the intrinsic $c$ parton inside the proton. In the former case, the lightest possible final states are $ \Lambda_c^+ D^0$.  If the photons are thermally distributed with a temperature of $T\sim (m_\Delta^2-m_p^2)/(2E_p)$, the rate of $p+\gamma \to D^0 +\Lambda_c^+$ will be suppressed by Bolzmann factor of $\exp [-E_\gamma/T]\sim\exp [-[(m_{D^0}+m_{\Lambda_c^+})^2-m_p^2)/(m_\Delta^2-m_p^2)]\sim 10^{-13}\ll 1 $, making the $\nu_\tau$ production completely negligible. The conclusion is robust even if the gas of photon has a modest Lorenz boost relative to the jet of the protons at the source. The $\nu_\tau$  production by the interaction of the photon with intrinsic $c$ quark inside the proton can be more significant. Studying this effect is however
  beyond the scope of the present paper. Our results applies  to the cosmogenic neutrinos produced by the scattering of the high energy protons off the thermally distributed CMB photon background. Notice that $\Delta^+ \to n \pi^+$ leads to $\pi^+ \to \mu^+ \nu_\mu$, $\mu^+ \to e^+ \nu_e \bar{\nu}_\mu$ so only $\nu_e$ (rather than a mixture of $\nu_e$ and $\bar{\nu}_e$) will be produced via the resonance $\Delta$ production. However, along the resonance production of $\Delta^+$,  $p \gamma \to p \pi^+ \pi^-$ and similar process can take place with a ratio of $O(1 \%)$ leading to the $\bar{\nu}_e$ production.
  
  For the case of the $p p$ scattering, the produced $\nu_\tau$ will also be sub-dominant but the ratio of the fluxes $F_{\nu_\tau}/F_{\nu_e}=2 F_{\nu_\tau}/F_{\nu_\mu} $ at the source can be still  larger than $10^{-3}$ (far exceeding the prediction for the sources with $p \gamma $ scattering) \cite{Abreu:2019yak}. Within the standard paradigm, $\nu_\tau$ can be produced en route by the oscillations of the $\nu_\mu$ and $\nu_e$ fluxes such that the flavor ratios at the detector become democratic. Thus, the possibility of a sub dominant $\nu_\tau$ component at the source is normally neglected. However, in the presence of new physics altering the standard three neutrino oscillation pattern, the sub dominant $\nu_\tau$
  component at the source may become important. In the following, we discuss two examples of such possibilities.
  
   {\section{$\nu_\tau$ mixed with a sterile neutrino, $\nu_s$ and the application for ANITA\label{anita+3+1}}
  While there are strong bounds on the mixing of $\nu_\mu$ and $\nu_e$ with $\nu_s$, the bound on the mixing between $\nu_\tau$ and $\nu_s$ is relatively relaxed. Indeed, for vanishing $U_{\mu 4}$ and $U_{e 4}$, $|U_{\tau 4}|^2$ can be as large as $0.18$ \cite{Dentler:2018sju}. If only $\nu_\tau$ mixes with $\nu_s$, in the absence of the $\nu_\tau$ production at the source, the flux will only contain $\nu_1$, $\nu_2$ and $\nu_3$, without any contribution from $\nu_4$. Let us take the flavor ratio at the source to be $F_e^s:F_\mu^s:F_\tau^s:F_s^s=1:2:f:0$ where $f=0$ for the $p\gamma$ source and $10^{-3} <f<1$ for the $pp$ source. After propagating the long distance between the source and the Earth, the mass eigenstates decohere. It is therefore more convenient to write the flux composition in the mass basis, $F_1:F_2:F_3:F_4=1:1:1+f(1-|U_{\tau 4}|^2):f|U_{\tau 4}|^2$. For
  ultrahigh energy neutrinos with $E_\nu \stackrel{>}{\sim}{\rm EeV}$, where active neutrinos become absorbed by the Earth, the sub-dominant $\nu_4$ component may have a significant impact.
  Ref. \cite{Cherry:2018rxj} has suggested $\nu_4$ with $|U_{\tau 4}|\sim 0.1$, as a solution for the anomalous $\nu_\tau$-like events observed by ANITA.  The mass of $\nu_4$ should be lighter than MeV; otherwise, it will decay into lighter neutrino states before traversing cosmological distances.
  Ref. \cite{Cherry:2018rxj} does not specify the source for $\nu_4$ but it argues that the source should be a transient such as  the flares of  AGNs, GRBs and TDE  to avoid the bounds from time integrated flux measured by ICECUBE and AUGER.\footnote{In Ref. \cite{Cherry:2018rxj},  the decay of ultraheavy dark matter (with a mass of few PeV) is suggested as the source of $\nu_4$ but such a source is not transient.}
  The basis of the idea in Ref. \cite{Cherry:2018rxj} is as follows. The cross section of the scattering of $\nu_4$ is given by that of $\nu_\tau$ times the mixing square [{\it i.e.,} $\sigma (\nu_4+{\rm nucleus})=|U_{\tau 4}|^2 \sigma (\nu_\tau +{\rm nucleus})$] so by taking $|U_{\tau 4}|^2 \sim 0.1$, the mean free path of $\nu_4$ will be of order of the chord size ({\it i.e.,} $L \sim 5000$ km) and the majority of the $\nu_4$ flux will survive traversing the Earth. Then, thanks to its  mixing with $\nu_\tau$ it can produce $\tau$ in the vicinity of the ANITA detector.  {This picture is not entirely valid because while the high energetic $\nu_4$ traverses the Earth, it loses its active component and will emerge in the vicinity of ANITA as $\nu_s$ rather than a mixture of $\nu_s$ and $\nu_\tau$, unable to produce $\tau$.} Let us study the propagation more systematically, considering a large mass splitting of $\Delta M^2$ between $\nu_4$ and the rest of the neutrino mass eigenstates, $\Delta M^2 \gg |\Delta m_{31}^2|$.
  
  The evolution of the neutrino states in the flavor basis, $\psi^T=(\nu_e,\ \nu_\mu,\ \nu_\tau,\ \nu_s)$ can be described by 
  \begin{eqnarray} i \frac{d \psi}{dt}&= &\left[U\cdot {\rm diag} (0,\Delta m_{21}^2/(2E_\nu),\Delta m_{31}^2/(2E_\nu), \Delta M^2/(2E_\nu) )\cdot U^\dagger\right.\cr &+& {\rm diag}(\sqrt{2} G_F(N_e-N_n/2),-\sqrt{2} G_FN_n/2,-\sqrt{2} G_FN_n/2,0)\cr
  &-& \left. i ~{\rm diag}(\Gamma/2,\Gamma/2,\Gamma/2,0)\right]\psi \end{eqnarray}
  where the only unusual term is the last one which takes care of the scattering of active neutrinos off matter. $\Gamma$ is the rate of the scattering which for $E_\nu \gg m_\tau$ is equal for all three active neutrinos. For these energies, $\Delta m_{21}^2 L/2E_\nu, \Delta m_{31}^2 L/2E_\nu \ll 1$ (with $L<2R_\oplus$) so we can safely set $\Delta m_{21}^2,\Delta m_{31}^2 \simeq 0$. We however keep $\Delta M^2$ and study the range for which it can lead to nontrivial effects.  Since we assume $U_{e4}=U_{\mu 4}=0$, we can write $U$ as 
  \begin{eqnarray}
  \left( \begin{matrix} U_{e1} & U_{e2} & U_{e3} & 0 \cr
  U_{\mu 1} & U_{\mu 2} & U_{\mu 3} & 0\cr
 \cos \alpha   U_{\tau 1} & \cos \alpha  U_{\tau 2} & \cos \alpha  U_{\tau 3} & \sin \alpha \cr
  - \sin\alpha   U_{\tau 1} & - \sin\alpha  U_{\tau 2} & -\sin \alpha  U_{\tau 3} & \cos \alpha \cr\end{matrix} \right),
  \end{eqnarray}
  where $U_{\beta i}$ with $\beta \in \{ e, \mu , \tau \}$ and $i \in \{ 1,2,3\}$ are the elements of the PMNS matrix. We can therefore write 
\begin{eqnarray}
|\nu_1\rangle &=&U_{e1}|\nu_e \rangle+ U_{\mu 1}|\nu_\mu \rangle +\cos \alpha U_{\tau 1}|\nu_\tau \rangle-\sin \alpha U_{\tau 1}|\nu_s \rangle\cr
|\nu_2\rangle &=&U_{e2}|\nu_e \rangle+ U_{\mu 2}|\nu_\mu \rangle +\cos \alpha U_{\tau 2}|\nu_\tau \rangle-\sin \alpha U_{\tau 2}|\nu_s \rangle\cr
|\nu_3\rangle &=&U_{e3}|\nu_e \rangle+ U_{\mu 3}|\nu_\mu \rangle +\cos \alpha U_{\tau 3}|\nu_\tau \rangle-\sin \alpha U_{\tau 3}|\nu_s \rangle\cr
|\nu_4\rangle &=&\sin \alpha |\nu_\tau\rangle +\cos \alpha |\nu_s\rangle.
   \end{eqnarray}
  Bearing in mind that $\Gamma L/2 \gg 1$, let us first discuss 
  the limit that $\Delta M^2 L/ 2E_\nu \ll 1$. In this limit, the $|\nu_i \rangle$ ($i \in \{1,2,3\}$)  states entering the Earth will emerge from the other side as 
  $|\nu_i \rangle \to -\sin \alpha U_{\tau i}|\nu_s\rangle$. In other words,  in the presence of mixing with the sterile neutrino, the three  light (mostly active) neutrino mass eigenstates will not be entirely absorbed and their small sterile component will survive. Similarly, $|\nu_4\rangle \to \cos \alpha |\nu_s\rangle$. Independently of starting with either of neutrino mass eigenstates, after traversing the Earth, only the $\nu_s$ component survives which cannot produce $\tau$ via the electroweak interactions. 
  In order to have the $\nu_s\to \nu_\tau$ oscillation, $1\stackrel{<}{\sim}
  \Delta M^2L/E_\nu$. Thus, taking $L\sim 5000$ km and $E_\nu\sim $EeV, $\sqrt{\Delta M^2}$ should be larger than $\sim 100$ eV. For $1\stackrel{<}{\sim}
  \Delta M^2L/E_\nu$, $P(\nu_4 \to \nu_\tau)$ can be as large as
  $\sim 0.1 \sin^2 \alpha$. Fig. \ref{dd} shows the probability of conversion of $\stackrel{(-)}{\nu}_3$ or $\stackrel{(-)}{\nu}_4$
arriving at the Earth into $\stackrel{(-)}{\nu}_\tau$ after traversing a chord of size $L$ in the mantle.  
  The difference between neutrino and antineutrino comes from flipping  the sign of the matter effects. The difference between the probabilities reflects the importance of matter effects. We have assumed  a mixing only between $\nu_s$ and $\nu_\tau$ parameters, $|U_{\tau 4}|^2=0.1$.
 We have focused on $E_\nu=$EeV to be close to the energies of the ANITA events. At this energy, the  cross section of the neutrino  is $1.1\times 10^{-32}$ cm$^2$ \cite{Safa,Cross-section}, leading to $\Gamma=0.003$ km$^{-1}$ for $\rho=4.5$ gr/cm$^3$ in the mantle. The oscillation probabilities are close to maximal around $\Delta M^2/2E_\nu=0.35 \Gamma\sim R_\oplus^{-1}$. As seen from the figure for these parameters $P(\bar{\nu}_3 \to \bar{\nu}_\tau)$ and  $P(\nu_4 \to \nu_\tau)$ can be as large as $\sim  10^{-3}$ and $0.02$, respectively. Indeed for $\Delta M^2/(2E_\nu)\sim \Gamma$, $P(\nu_4 \to \nu_\tau)\sim 0.1 \sin^2 \alpha$. Moreover, as long as $\sin \alpha >0.01$, $P(\nu_3 \to \nu_\tau) \sim 0.1 \sin^4 \alpha$. For smaller $\sin \alpha$, the effect of the oscillation between $\nu_s$ and $\nu_\tau$ at energies of EeV in the Earth is negligible so the $\sin^4 \alpha$ scaling does not hold.
 Taking   $\Delta M^2/E_\nu \gg R_\oplus^{-1}$, we find  that $P(\nu_4 \to \nu_\tau)\simeq P(\bar{\nu}_4 \to \bar{\nu}_\tau)\sim 0.01 (\sin^2 \alpha/0.1)$ and $P(\nu_3 \to \nu_\tau)\simeq P(\bar{\nu}_3 \to \bar{\nu}_\tau)\to 0$.
 
  %%%%%%%%%%%%%%%%%%%%%%%%%%%%%%% 
  \begin{figure}[h]
  	\hspace{0cm}
  	\includegraphics[width=0.60\textwidth, height=0.5\textwidth]{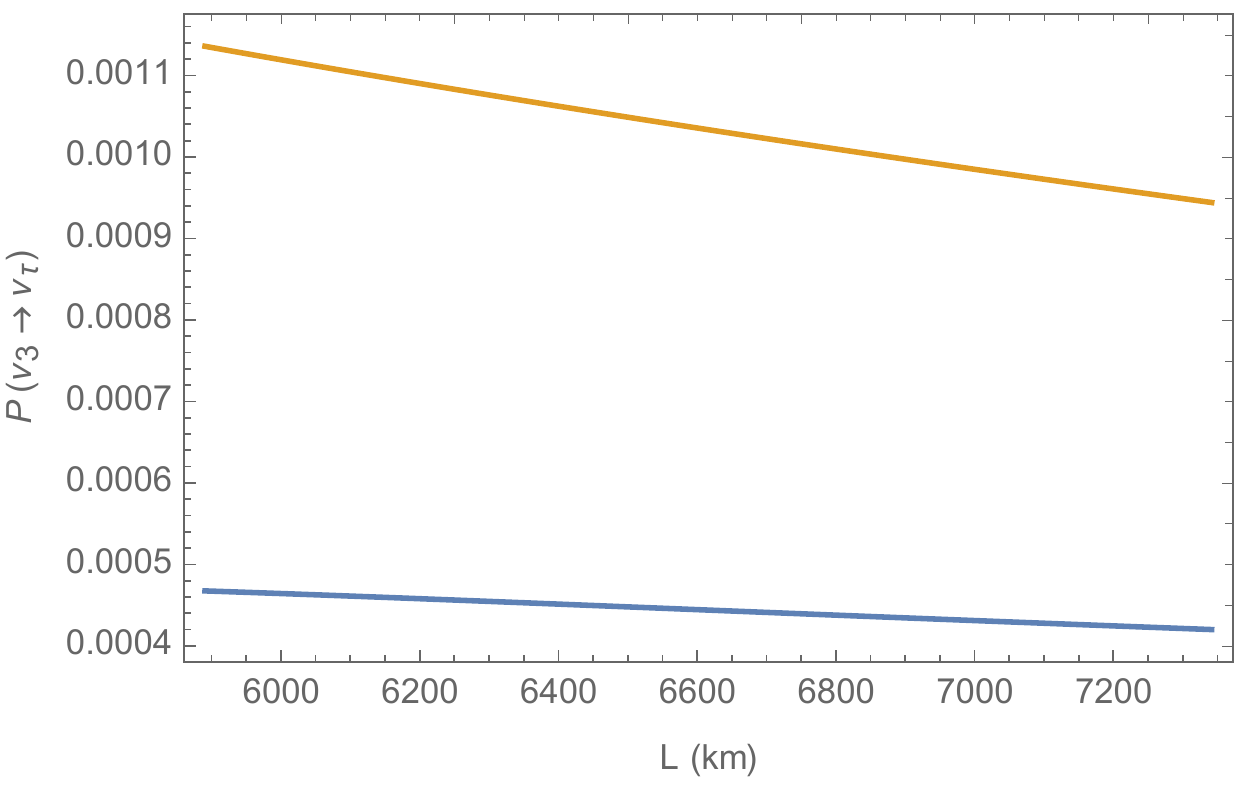}
  	\includegraphics[width=0.60\textwidth, height=0.5\textwidth]{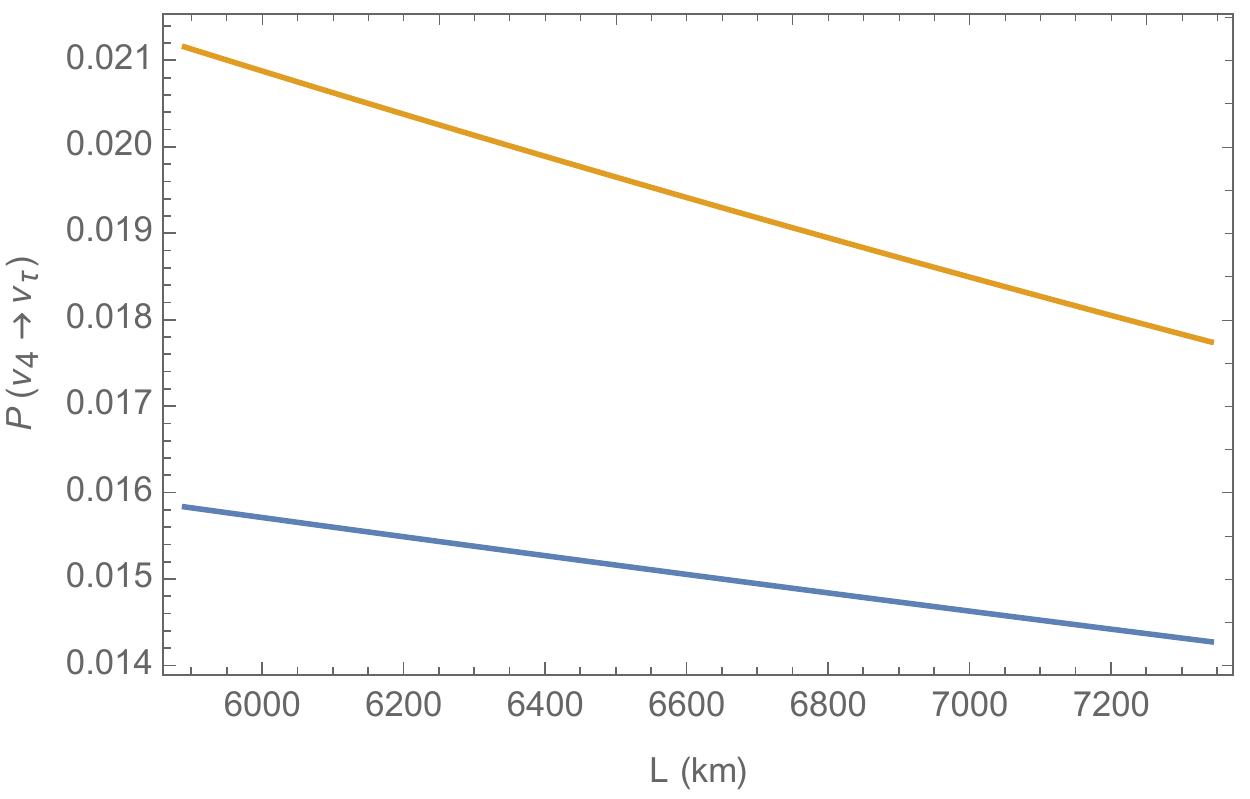}
  	\caption[...]{ Oscillation probability versus the size of the traversed chord. We have taken $E_\nu=$EeV, $\Gamma = 0.003~{\rm km}^{-1}$ (corresponding to $\rho=4.5$ gr/cm$^3$ and  cross section of $1.1\times 10^{-32}~{\rm cm}^2$ taken from \cite{Safa}), $|U_{\tau 4}|^2=\sin^2 \alpha=0.1$ and $\Delta M^2/2 E_\nu=0.35 \Gamma$ which corresponds to $\sqrt{\Delta M^2}=643$~eV. The orange and blue lines respectively correspond to the antineutrino and neutrino modes.
  	}
  	\label{dd}
  \end{figure} 
 Consider a neutrino flux with a flavor ratio of $(F_{\nu_e}:F_{\nu_\mu}:F_{\nu_\tau}:F_{\nu_s})=(1:2:f:0)$ which, as discussed before, at arrival at the Earth can be described in the mass basis as $(F_{1}:F_{2}:F_{3}:F_{4})=(1:1:1+f(1-|U_{\tau 4}|^2):f|U_{\tau 4}|^2)$ with a normalization of $F$ at arrival. As shown in \cite{Safa,Pizzuto:2019vyj}, $\nu_\tau$ flux at this energy leads to a regenerated lower energy  neutrino flux on which ICECUBE has put a bound. The bound on $F$ for EeV energies is $10^{-7}~{\rm cm}^{-2}$.
 The contributions from $\nu_3$ and $\nu_4$ to the $\nu_\tau$ events at the detector will be respectively given by 
 $F(1+f)P(\nu_3 \to \nu_\tau)\sim 0.1 F(1+f)\sin^4 \alpha$ and $Ff\sin^2\alpha P(\nu_4 \to \nu_\tau)$. As discussed before, the contribution from the $\nu_\tau$ production at the source, $f$, is smaller than 1. Thus, as long as $\nu_s$ is not directly produced at the source, $\nu_3 \to \nu_\tau$ will dominate the $\tau$ events.
 
 Considering the upper bound on $F$, even in the favorable parameter range $\sin^2 \alpha =0.1$ and $\Delta M^2/(2 E_\nu)\sim 0.3 \Gamma$, the flux of $\nu_\tau$ after passing the mantle cannot be larger than $10^{-10}$ cm$^{-2}$. Considering that the acceptance of ANITA is $\sim 2 \times 10^9$ cm$^2$ \cite{Safa} (See also \cite{Romero-Wolf:2018zxt}), this means the number of $\tau$ like events will be too small to explain the  ANITA events. Some  possibilities to enhance the predicted number of $\tau$ events at ANITA are the following:
 1) If $\nu_s$ also mixes with $\nu_e$ or $\nu_\mu$, the flux of $\nu_4$ reaching the Earth will be given by $F(2|U_{\mu 4}|^2+|U_{e 4}|^2)$. The upper bound on $|U_{\mu 4}|^2$ and $|U_{e 4}|^2$ are of order of $10^{-2}$ \cite{Dentler:2018sju}. The flux after traversing the Earth will then be $F(2|U_{\mu 4}|^2+|U_{e 4}|^2) P(\nu_4 \to \nu_\tau) \stackrel{<}{\sim} ({\rm few }\times 10^{-4}) F$. Taking $F\sim 10^{-7}$ cm$^{-2}$, again the number of events at ANITA will be smaller than 0.1.
 2) Production of $\nu_s (\simeq \nu_4)$ at the source such that $P(\nu_4\to \nu_\tau )F_{\nu_s}/F_{\nu_\mu}\stackrel{>}{\sim} 0.02$. That is the flux of $\nu_s$ at the source should be larger than the flux of $\nu_\tau$. 
 Obtaining such large $\nu_s$ flux at the source is very challenging, especially considering that there are strong bounds on the branching ratios of meson decays into $\nu_s$: $Br(\pi^+ \to e^+\nu_s)\ll Br(\pi^+ \to \mu^+\nu_s)<10^{-3}$ and $Br(\pi^0 \to \bar{\nu}\nu_s)<10^{-7}$.
 3) Relaxing the upper bound on $F$. The upper bound comes from the regeneration of the lower neutrino flux from high energy $\nu_\tau$ flux. 
 The  $\tau$ 
   produced from the CC interaction of $\nu_\tau$ decays back into $\nu_\tau$ before losing a significant portion of its energy but the muon produced by the 
 CC interaction of $\nu_\mu$ loses significant energy before decaying back into $\nu_\mu$. The electron from CC interaction of $\nu_e$ becomes of course absorbed before producing $\nu_e$. If there is a mechanism like the one introduced in \cite{Farzan:2018pnk}, to maintain $\nu_e$ and $\nu_\mu$ as the initial flavor, there will be no reproduced lower energy flux to be detected by ICECUBE so the bound on $F$ can be relaxed. We will return to this possibility after revisiting the bounds on the scenario proposed in \cite{Farzan:2018pnk}.

  In future with detectors with larger acceptance such as POEMMA  \cite{Olinto:2019euf} there will be interesting possibility of testing this scenario with $\Delta M^2/(2 E_\nu)\sim (0.1-1)\Gamma$ and a value of $\sin^2 \alpha$ close to the present bound, provided that $F$ is also close to the ICECUBE bound. In this exciting situation, three spectacular observations will be in the corner: (1)
Further data from ICECUBE can observe transient  neutrino fluxes with energies of few PeV produced by regeneration in the Earth \cite{Safa}. (2)   
A coincident transient signal with energy of EeV observed by the successors of ANITA  coming from deep down the Earth after crossing mantle and/or the core with directionality matched with the accompanying lower energy regenerated neutrino flux. (3) The measurement of nonzero $|U_{\tau 4}|$ in the experiments such as the upgraded FASER$\nu$ detector during high-luminosity LHC \cite{Abreu:2019yak} or the improvement of the bounds on the unitarity of the PMNS matrix.

  \section{Interaction with ultralight dark matter\label{UDM}}
  In Ref. \cite{Farzan:2018pnk}, we have proposed a model for the interaction of the neutrinos  with the background
  ultralight dark matter field which behaves like a classic complex scalar. The model that we have proposed is based on a $L_\mu-L_\tau$ gauge symmetry. After integrating out the intermediate gauge boson, the interaction can be written as the following dimension six operator:
  \be \label{D6} \frac{i g_\alpha}{\Lambda^2}(\phi^\dagger \partial_\mu \phi-\phi \partial_\mu \phi^*)(\bar{\nu}_\alpha \gamma^\mu \nu_\alpha).\ee
  Such an interaction will act as an effective mass of form 
  \be V_\alpha \nu_\alpha^\dagger \nu_\alpha \ \ \ \ {\rm with}
  \ \ \ \ V_\alpha=\frac{\rho_{DM}}{m_{DM}}\frac{g_\alpha}{\Lambda^2}.\ee
  The density of the Dark Matter (DM) in the DM halos can be $10^5-10^6$ times the average DM density in the universe. Consider  a neutrino with an energy of $E_\nu$ being produced in a source inside the DM halo of a galaxy and detected by a detector in the solar system which is of course again immersed in the DM halo. Both at the position of the source and at that of the detector, $V_\alpha$ can be much larger than $\Delta M^2/(2E_\nu)$ which implies that at the source and at the detector, the flavor eigenstates and the effective mass (energy) eigenstates coincide. Considering that the DM distribution is smooth, the transition will be adiabatic, implying that the flavor of the neutrinos will not change in propagation from the source to the detector.
  Details are elaborated in Ref.  \cite{Farzan:2018pnk} so will not be repeated here. The bottom line is that the flavor ratio $F_{\nu_e}:F_{\nu_\mu}:F_{\nu_\tau}=(1,2,0)$ will be maintained up to the detector and unlike the SM prediction, $\nu_\tau$ will not be produced via oscillation. Discovery of the $\nu_\tau$ events by ICECUBE confirms the SM picture up to energies of a few PeV so 
  \be \label{bound} \frac{g_\alpha}{\Lambda^2}\frac{\rho_{DM}}{m_{DM}}\ll \frac{\Delta m_{31}^2}{2E_\nu}\sim 10^{-18}~e\rm{V}.\ee
  Taking $({g_\alpha}/{\Lambda^2})({\rho_{DM}}/{m_{DM}})\sim 10^{-19}~e\rm{V}$, the cosmic neutrinos with energies lower than 10 PeV detected so far by ICECUBE will not be affected but higher energy neutrinos,  which can be detected by the next generation telescopes such as  GRAND \cite{Fang:2017mhl} or POEMMA \cite{Olinto:2019euf}, can be affected. In the following, we discuss different possible sources and the strategies to unravel the secrets of DM by studying the messengers from them.
  
  A guaranteed example for EeV neutrinos is the GZK or cosmogenic neutrinos produced by the scattering of the cosmic ray off photons of the CMB. Considering that the CMB is distributed uniformly all over the universe without clustering in any halo,
  the mean free path of protons with energy exceeding $5\times 10^{19}~e$V is 6 Mpc \cite{Ruffini:2015oha} which is larger than the halo size. That is even if the cosmic ray originates in a halo with large $\rho_{DM}$, the production of the neutrinos will take place outside the halo where $\rho_{DM}$ is smaller than $\sim 10^{-5}$ times $\rho_{DM}$ in the halo. Combining this with the recent bound from observation of the tau neutrino events ({\it i.e.,} Eq. \ref{bound}), we conclude  that  the dark matter effects for the oscillation of the GZK neutrino are suppressed until they reach the Milky Way halo (or the DM halos en route).
  When  neutrino flux reaches the halo, it can be described as mass eigenstates with the ratio $F_{\nu_1}:F_{\nu_2}:F_{\nu_3}=1:1:1$. Due to DM effects $\nu_1$, $\nu_2$ and $\nu_3$ will convert to coherent states $\nu_e$, $\nu_\mu$ and $\nu_\tau$, respectively but we cannot observationally distinguish this situation with pure SM without DM effects. Even if the neutrinos pass through the halo of another galaxy, the prediction does not change because the neutrinos exiting the halo will be in form of mass eigenstates with again a democratic ratios. DM effects become significant  only when both the source and the detector are immersed in the regions with large $\rho_{DM}$. Of course, this statement applies for the case when the source is also inside the Milky Way dark matter halo \cite{deSalas:2016svi}.
  
  The three upcoming key players in exploring the cosmic neutrino flux will be ARA \cite{ARA}, ARIANNA \cite{ARIANNA}, POEMMA \cite{Olinto:2019euf} and GRAND \cite{Fang:2017mhl}. ARA and ARIANNA will probe the  Antarctic ice and like ANITA, they will be sensitive to the Askaryan radiation from all three neutrino flavors. Moreover, they will be sensitive to  Extensive Air Shower (EAS) from hadronic  decay of $\tau$ produced by  the Charged Current (CC)  interaction of $\nu_\tau$. GRAND will be only sensitive  to the $\nu_\tau$ events.
  
  Taking a democratic flavor ratio for the neutrinos, the sensitivity of the first phase of GRAND after 3 years of data taking, covering an area of 10000 km$^2$ will be comparable to that of projected 3 year reach of ARA and ARIANNA \cite{Fang:2017mhl}.  If the latter detectors discover a significant neutrino flux by detecting Askaryan  radiation but no $\nu_\tau$ counterpart is observed, it will be hint for (1) an interaction of type with $\frac{g_\alpha }{\Lambda^2}   \frac{\rho_{DM}}{m_{DM}}\sim 10^{-20}-10^{-19}$ eV;
   (2) the flux is originated from halos where $\rho_{DM}$ is large. On the other hand, if these detectors confirm the (close to) democratic flavor ratio, we cannot simply constrain $g_\alpha /\Lambda^2$ because, the origin of the flux may be outside the halo ({\it e.g.,} cosmogenic neutrinos). For the time being, the uncertainty in the prediction of the neutrino flux is as large as two orders of magnitude, ranging from just above the sensitivity of the phase I of GRAND after 3 years to the sensitivity of an upgrade of GRAND with  200000 km$^2$ coverage after three years \cite{Fang:2017mhl}. The uncertainty comes from the uncertainty in the composition, spectra and source and the evolution history of the cosmic ray. If these uncertainties are reduced and the cosmogenic neutrino flux is predicted to be below the sensitivity of three years  data of these experiments, the detection of a cosmic neutrino flux with energies of EeV would indicate a source such as an AGN in a halo. The detection of  $\nu_\tau$ events by GRAND and/or ARIANNA \cite{ARIANNA} and ARA \cite{ARA} will then improve the upper bound on  $(g_\alpha/\Lambda^2)(\rho_{DM}/m_{DM})$ derived from the observation of $\nu_\tau$ events by ICECUBE by three orders of magnitudes.
   
   In the above discussion, we assumed time integration over the diffuse fluxes of neutrinos. In case of multimessenger events (composed of electromagnetic signal plus neutrinos) 
   from transient activities of the point sources, we can make sure that the neutrino flux originates from a region with large enough $\rho_{DM}$. Of course, the number of events from a single outburst will be too small to test the $\nu_e:\nu_\mu:\nu_\tau=1:1:1$ hypothesis. In the lucky situation of observing a significant number of such multimessenger events, invaluable information can however be derived on $(g_\alpha/\Lambda^2)(\rho_{DM}/m_{DM})$.

   \section{Revisiting the 3+1 scheme in the background of ultralight DM\label{Shole}} Let us study the effects of the interaction of type shown in equation (\ref{D6}) on the evolution of the neutrino flavor within the 3+1 neutrino mass scheme. Taking $(g_\alpha/\Lambda^2)(\rho_{DM}/m_{DM})|_{\rm halo}\sim 10^{-19}$ eV, if the mass of the sterile neutrinos is larger than 1~eV, the effect of dark matter halo on the effective mixing between $\nu_s$ and active neutrinos will be negligible. However, in the early universe before neutrino decoupling,
   $(g_\alpha/\Lambda^2)(\rho_{DM}/m_{DM})|_T \sim 10^{5}~{\rm eV} (T/{\rm MeV})^3$. The effective  mixing between active and sterile neutrinos will therefore be suppressed in the early universe as long as the sterile neutrino mass is below few 100 keV. As a result, the production of the sterile neutrinos in the early universe will be suppressed so the bounds from cosmology on the 3+1 scheme can be avoided, saving the sterile neutrino solution to the  LSND and MiniBooNE anomalies and reactor neutrino deficit \cite{Farzan:2018pnk}.
   
   Let us now revisit the 3+1 solution to the ANITA anomalous events which we discussed in sect \ref{anita+3+1} in the presence of ultralight DM effects. If $\nu_e$ and $\nu_\mu$ do not mix with $\nu_s$, their evolution will be similar to what described in Ref. \cite{Farzan:2018pnk} and reviewed  in sect. \ref{UDM}: $\nu_e$ and $\nu_\mu$ will maintain their flavor from a source in a DM halo up to reaching the Earth, they will then become absorbed in the Earth without regenerating $\nu_\tau$ in the lower energies so the bound in \cite{Safa} does not apply but the ANITA events cannot be explained, either. Let us however consider the cases that $\nu_s$ simultaneously  mixes with $\nu_\tau$ and $\nu_\mu/\nu_e$. Since the bound on the mixing of $\nu_\mu$ and $\nu_s$ is relatively strong  \cite{Dentler:2018sju}, we will only consider the mixing of $\nu_s$ with $\nu_\tau$ and $\nu_e$  with mixing angles $\alpha$ and $\beta$, respectively. Taking the sterile neutrino mass $M\sim 100$ eV-keV, $\nu_e$ can oscillate to $\nu_\tau$ and $\nu_s$ en route. Starting with a flavor composition of $F_{\nu_e}:F_{\nu_\mu}:F_{\nu_\tau}:F_{\nu_s}=1:2:0:0$, the flux at the detector in the effective mass basis can be  described as $(F_{\tilde{1}}:F_{\tilde{2}}:F_{\tilde{3}}:F_{\tilde{4}})=(\cos^2 \beta:2:\sin^2 \alpha \sin^2 \beta:\cos^2 \alpha \sin^2 \beta)$ where $|\tilde{1}\rangle =\cos \beta |\nu_e\rangle -\sin \beta |\nu_s\rangle$, $|\tilde{2}\rangle = |\nu_\mu\rangle$, $|\tilde{3}\rangle =-\sin \alpha\sin \beta |\nu_e\rangle +\cos \alpha |\nu_\tau \rangle -\sin\alpha \cos \beta |\nu_s\rangle$ and $|\tilde{4}\rangle =\cos \alpha\sin \beta |\nu_e\rangle +\sin \alpha |\nu_\tau \rangle +\cos\alpha \cos \beta |\nu_s\rangle$. Although because of the DM effects through the interaction in Eq (\ref{D6}), the $\Delta m_{31}^2$ and $\Delta m_{21}^2$ modes of the $\nu_e \to \nu_\tau$ oscillation are suppressed, because of the $\Delta m_{41}^2=\Delta M^2$ splitting, $\nu_e$ can partially convert into $\nu_\tau$.  The probability of $\nu_e\to \nu_\tau$ from a source in a halo up to the surface of the Earth can be written as 
   \be \label{prob} P(\nu_e \to \nu_\tau)|_{ surface}^{\oplus}=2 |U_{\tau 4}|^2|U_{e 4}|^2=4 \times 10^{-3}\frac{|U_{\tau 4}|^2}{0.1}\frac{|U_{e 4}|^2}{0.02}.\ee
    The $\nu_e$ flux produced in a source at a halo can be decomposed as $\nu_e=\cos \beta\tilde{\nu}_1-\sin \alpha \sin \beta \tilde{\nu}_3 + \cos \alpha \sin \beta \tilde{\nu}_4$.  Travelling the long distance between the source and the Earth, the $\tilde{\nu}_1$, $\tilde{\nu}_3$  and $\tilde{\nu}_4$ components decohere. The probability of $\nu_e \to \nu_\tau$ after crossing the Earth will be given by
   \be P(\nu_e \to \nu_\tau)|_{ cross}^{\oplus}=\cos^2 \beta P(\tilde{\nu}_1 \to \nu_\tau)+\sin^2 \alpha \sin^2\beta P(\tilde{\nu}_3 \to \nu_\tau)+\cos^2 \alpha \sin^2\beta P(\tilde{\nu}_4 \to \nu_\tau),\ee
   where $P(\tilde{\nu}_i \to \nu_\tau)$ takes care of the Earth matter effects, absorption in the Earth and the $\Delta M^2$ oscillation modes.  Fig. (\ref{nd}) shows $P(\nu_e \to \nu_\tau)$ and $P(\bar{\nu}_e \to \bar{\nu}_\tau)$  from a source in a halo up to the detector after crossing a chord of size $L$. To draw the figure, we have taken $|U_{\tau 4}|^2 \simeq \sin^2 \alpha=0.1$ and 
  $|U_{e 4}|^2 \simeq \sin^2 \beta=0.02$ which satisfy the present bounds \cite{Dentler:2018sju}. The rest of the input is similar to what is assumed to draw Fig. \ref{dd}. Comparing Fig. \ref{nd} with Eq. (\ref{prob}), we find that the ratio of the $\nu_\tau$ flux at the arrival on the Earth to that after traversing chords of size 5000~km-7500~km will not be larger than $\sim 10$. As seen in Fig \ref{nd}, 
  in this model $\nu_\tau$ with EeV energy can emerge and therefore scatter to $\tau$ deep inside the Earth so the spectral shape of the regenerated neutrino flux at lower energies ({\it i.e.,} the flux from $\nu_\tau \to \tau \to \nu_\tau\to ....\to \nu_\tau$) will be different from that predicted within the SM \cite{Safa}. Computing the spectral shape is beyond the scope of the present paper but we can estimate the total lower energy regenerated $\nu_\tau$ flux  as 
  $$ F \Gamma \int_0^L P(\nu_e \to \nu_\tau)|^\oplus_{cross}[x] dx \sim 0.01 F,$$
  where $F$ is the time integrated  flux of $\tilde{\nu}_1\simeq \nu_e$ at the surface and $L\sim 5000-7500$ km is the chord size. Within the SM, the $\nu_\tau$ flux at the surface  would be equal to the $
 \nu_e$ flux, $F$ which would be entirely absorbed and  would lead to regenerated lower energy $\nu_\tau$ flux. Thus, the bound of $10^{-7}$ cm$^{-2}$ derived in \cite{Safa} on the $\nu_\tau$ flux within the SM framework  should be reinterpreted as $0.01 F<10^{-7}$~cm$^{-2}$ or $F<10^{-5}$ cm$^{-2}$ within the present model. Saturating this bound, we expect
  \be 
  F P(\nu_e \to \nu_\tau)|_{cross}^\oplus
Acc=10 \frac{F}{10^{-5}~{\rm cm}^{-2}}\frac{P(\nu_e\to\nu_\tau)}{0.0005}  \frac{Acc}{2\times 10^9 ~{\rm cm}^2},\ee
where $Acc$ is the acceptance of the neutrino telescope. Taking $Acc$ equal to the acceptence of  ANITA, we find that the number of events can be as large as 10. Thus the ANITA events can readily be explained with $F\sim 10^{-6}$~cm$^{-2}$.

  \begin{figure}[h]
  	\hspace{0cm}
  	\includegraphics[width=0.70\textwidth, height=0.6\textwidth]{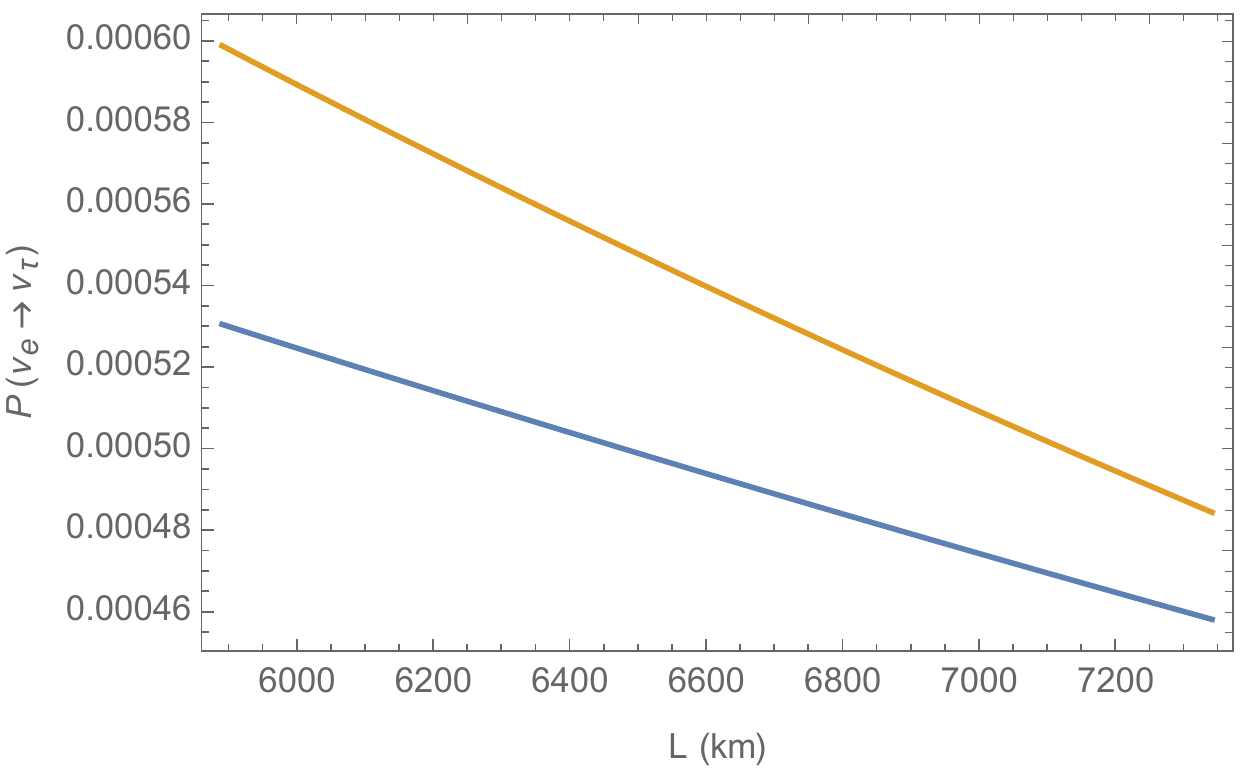}
  	\caption[...]{ Probability  $\nu_e \to \nu_\tau$ in the presence of DM effects within the $3+1$ scheme versus the size of the traversed chord. We have assumed that both in the solar system and in the production site, the dark matter density is relatively large with $V_\alpha \gg \Delta m_{31}^2/(2E_\nu)$ but still $V_\alpha \ll \Delta M^2/(2 E_\nu)$ so the effective mixing between active and sterile neutrinos remain equal to the values in the vacuum but the effective mixings between active neutrinos become suppressed.  We have taken $|U_{\tau 4}|^2=0.1$ and $|U_{e4}|^2=0.02$. The rest of the parameters are as described in the caption of Fig. (\ref{dd}). The orange and blue lines respectively correspond to the antineutrino and neutrino modes.
  	}
  	\label{nd}
  \end{figure}
  Let us discuss the possible source for the ANITA events. To produce $\nu$ with an energy of EeV, the energy of proton impinging on the background photon and proton should be around 10 EeV. To have the $p+\gamma \to \Delta$ resonance, $\gamma$ has to have an energy of order of 0.05~eV. For a thermal distribution of photon with a temperature of 0.05~eV or larger, taking the cross section equal to 120 $\mu b$ \cite{Dermer:2006bb}, the mean free path of the proton will be of order of the size of the jets of AGNs. In case of proton proton interaction, the largest contribution comes from a parton in the incident proton carrying momentum fraction of 0.1 scattering on a very low momentum fraction parton such that their center of mass energy is of order of  1 GeV or smaller. However, partons in other ranges can also produce $\nu_\tau$ (reaching the Earth surface) as well as $\nu_e$ and $\nu_\mu$ at energies of order of $\sim$PeV.
  % Absence of an accompanying signal  at ICECUBE associated with ANITA signals therefore rules out the $p-p$ scattering source.

    Let us entertain the possibility  of explaining the two $\nu_\tau$ events observed by ICECUBE within this scenario, ruling in $V_\alpha \sim 10^{-18}$ eV. For relatively small values of $\Delta M^2$ with $E_\nu/\Delta M^2\gg R_{\oplus}$, we then expect $F_{\nu_\tau}/F_{\nu_e}=2 |U_{e 4}|^2|U_{\tau 4}|^2<4\times 10^{-3}$ which is ruled out by ICECUBE. For larger $\Delta M^2$ with $E_\nu/\Delta M^2\sim R_{\oplus}$, this ratio will still remain smaller than few percent which is again ruled out.
    
    \section{Conclusions \label{Fin}}
    In the next decades, arrays of radio telescopes with huge coverage will make a breakthrough in studying the anticipated cosmogenic neutrinos as well as other possible neutrino fluxes with EeV range energies. While GRAND detector will be sensitive to the $\nu_\tau$ flux by observing Extensive Air Showers initiated by $\nu_\tau$, ARA and ARIANNA can both detect such showers from $\nu_\tau$ as well as the Askaryan radiation from all  flavors. It seems that from observational point of view in this extremely high energies, the $\tau$ flavor holds a special place amongst the flavors. If new physics leads to a significant deviation from the canonical democratic flavor composition, it would imply drastic consequences  for the discoveries to be made by these upcoming detectors. 
    We have focused on the following beyond SM scenarios that can have non-trivial effects on the flavor composition of cosmic neutrinos:
    (1) The $3+1$ neutrino scheme with $U_{\tau 4}\ne 0$ which is motivated by the two anomalous $\nu_\tau$ events reported by ANITA; (2) Current-current interaction between neutrinos and the background ultralight Dark Matter (DM); (3) A combination of (1) and (2).
    
    We have discussed and corrected a misconception that existed in the literature about the propagation of a coherent linear combination of sterile and active neutrino states in matter in the presence of CC scattering and absorption.
    We have shown that at energies of EeV when the Earth becomes opaque for the active neutrinos, the active components of mass eigenstates become absorbed. As a result, for $\Delta M^2 R_\oplus /E_\nu\ll 1$,  $\nu_4$ will convert to $\nu_s$ after crossing chords with size larger than $\sim 1000$ km. As a result, high energy $\nu_4$ crossing the Earth could not produce the $\tau$ events detected by ANITA. We have however shown that if the  length of the oscillation between active and sterile neutrino is comparable to the chord size, $P(\nu_4\to \nu_\tau)\sim 0.01$ and $P(\nu_3\to \nu_\tau)\sim 10^{-3}$. For $\Delta M^2 R_\oplus /E_\nu \gg 1$, $P(\nu_4 \to \nu_\tau)$ remains of order of $O(1 \%)$ but $P(\nu_3\to \nu_\tau)$ vanishes. Thus, to have relatively large $P(\nu_3\to \nu_\tau)$ and $P(\nu_4\to \nu_\tau)$, the mass of the fourth neutrino should be $O(500~ e{\rm V})$. On the other hand, the mass of fourth neutrino has to be smaller than MeV to make its lifetime long enough to travel cosmological distances \cite{Cherry:2018rxj}. With such values for the parameters, the fourth neutrino will be produced in the early Universe and can contribute to extra relativistic degrees of freedom  on which there are strong bounds from CMB and big bang nucleosynthesis. We have also shown that within the $3+1$  solution to the ANITA anomalous events a large flux of $\nu_\tau$ is predicted to enter the Earth that will reproduce lower energy (PeV) neutrino flux via $\nu_\tau \to \tau ...\to \nu_\tau$ in the Earth detectable by ICECUBE. The corresponding bound discussed in \cite{Safa} will rule out this solution. We have argued that turning on other mixing angles ({\it e.g.,} $U_{e 4}$ and $U_{\mu 4}$) or exotic decay modes for meson at source of form $\pi \to \mu \nu_s$ or $e\nu_s$ cannot be a remedy because of the already existent bounds on the mixing parameters and/or on the branching ratios of the exotic modes. 
    
    We have then discussed the implications of an interaction of form (\ref{D6}) between ultralight dark matter  and neutrinos for the flux of extremely high energy neutrinos with energies of EeV. We have interpreted the recent observation of the 100 TeV and 2 PeV $\nu_\tau$ events in the ICECUBE data \cite{Abbasi:2020zmr} as an upper bound on the coupling of this interaction. Despite this rather tight bound, the effect can be still significant for higher energy neutrinos detectable by ANITA and its forthcoming successors. We have argued that although the dark matter effects can be significant for the propagation of each flavor of cosmogenic neutrinos when they enter the DM halo of our galaxy,  the DM effect will not alter the canonic $\nu_e:\nu_\mu:\nu_\tau=1:1:1$ prediction for the  cosmogenic neutrinos.
    The reason is that the cosmogenic neutrinos are produced in regions where DM density is low.
     However, if the source is located in a halo (either the Milky Way DM halo or the halo of another galaxy), the DM effect may maintain the original flavor ratio of $\nu_e:\nu_\mu:\nu_\tau=1:2:0$ up to the Earth. This means while the ARA and ARIANNA detectors will observe 
     the Askaryan effects from the $\nu_e$ and $\nu_\mu$ fluxes, GRAND or other detectors will not observe Extensive Air Shower signals from the $\nu_\tau$ flux. We have argued that by improving the uncertainties on the prediction of the total flux of cosmogenic neutrinos, the ratio $\nu_\tau /(\nu_e+\nu_\mu)$ derived from combined analysis of the data from  these future detectors can yield information on the coupling between neutrinos and dark matter of form (\ref{D6}).
    
    We have then studied the $3+1$ scheme in the presence of an interaction of form (\ref{D6}) with ultralight dark matter. As first proposed in \cite{Farzan:2019yvo}, this form of interaction can save the $3+1$ scheme from cosmological bounds. Moreover, in the presence of this interaction, the conversion of  cosmic $\nu_e$ ($\nu_\mu$)
  to $\nu_\tau$ will be suppressed by $|U_{e 4}|^2|U_{\tau 4}|^2$  ($|U_{\mu 4}|^2|U_{\tau 4}|^2$). This helps tp relax the bound from non-observation of accompanying regenerated lower energy flux by ICECUBE. We have shown that with $U_{e 4}$ and $ U_{\tau 4}$ saturating the present bound, the ANITA events find a reasonable solution within the $3+1$ scheme with an interaction of form  (\ref{D6}).
  
  %%%%%%%%%%%%%%%%%%%%%%%%%%%%%%%
  %%%%%%%%%%%%%%%%%%%%%%%%%%%%%
    
  \section*{Acknowledgment} The author is very grateful to A. Smirnov, A. Kheirandish and S. Palomares-Ruiz  for very useful comments and the encouragement. She also thanks K. Murase for the useful comments.
  This project has received funding /support from the European Union’s Horizon 2020 research and innovation programme under the Marie Skłodowska -Curie grant agreement No 860881-HIDDeN. The author has received partial financial support from Saramadan under contract No.~ISEF/M/98223 and No.~ISEF/M/99169.
  %%%%%%%%%%%%%%%%%%%%%%%%%%%%%%%%%
  %%%%%%%%%%%%%%%%%%%%%%%%%%%%%%

\end{document}